\documentclass[aps,prl,twocolumn,preprintnumbers,amsmath,amssymb,superscriptaddress]{revtex4}%

\usepackage{graphicx}%
\usepackage{dcolumn}
\usepackage{amsmath}
\usepackage{color}
\usepackage{multirow}

\begin{document}


\title{Pressure-induced magnetism in iron-based superconductors \\ $A$Fe$_2$As$_2$ ($A=$ K, Cs, Rb)}

\author{Rustem Khasanov}
 \email[Corresponding author: ]{rustem.khasanov@psi.ch}
\author{Zurab Guguchia}
\author{Elvezio Morenzoni}
\author{Chris Baines}
 \affiliation{Laboratory for Muon Spin Spectroscopy, Paul Scherrer
Institute, CH-5232 Villigen PSI, Switzerland}
\author{Aifeng Wang}
\author{Xianhui Chen}
 \affiliation{Hefei National Laboratory for Physical Sciences at Microscale and Department of Physics, University of Science and Technology of China, Hefei, Anhui 230026, China}
\author{Zbigniew Bukowski}
 \affiliation{Institute of Low Temperature and Structure Research, Polish Academy of Sciences, 50-422 Wroclaw, Poland}
\author{Fazel Tafti}
 \affiliation{Department of Physics, Boston College, Chestnut Hill, Massachusetts 02467, USA}
 \affiliation{D\'{e}partement de physique \& RQMP, Universit\'{e} de Sherbrooke, Sherbrooke, Qu\'{e}bec J1K 2R1, Canada}

\begin{abstract}
The magnetic properties of iron-based superconductors $A$Fe$_2$As$_2$ ($A=$K, Cs, and Rb), which are characterized by the V-shaped dependence of the critical temperature ($T_{\rm c}$) on pressure ($P$) were studied by means of the muon spin rotation/relaxation technique.
In all three systems studied the magnetism was found to appear for pressures slightly below the critical one ($P_{\rm c}$), {\it i.e.} at pressure where $T_{\rm c}(P)$ changes the slope. Rather than competing, magnetism and superconductivity in $A$Fe$_2$As$_2$ are coexisting at $P\gtrsim P_{\rm c}$ pressure region.
Our results support the scenario of a transition from one pairing state to another, with different symmetries on either side of $P_{\rm c}$.
\end{abstract}
\maketitle


Since the discovery of iron-based superconductors, much effort was devoted to identify the pairing mechanism responsible for their high critical temperature \cite{Johnston_AIP_2010, Chubukov_ARCMP_2012}. While some properties of iron-based superconductors are reminiscent of the cuprate superconductors, the differences between the two families are considerable. Iron-based superconductors (Fe-SC's) are generally believed to have $s$-wave pairing symmetry. They are multi-band superconductors, with an energy gap that varies significantly amongst the various Fermi surface sheets \cite{Evtushinsky_NJP_2009}. Their gap structure is not universal and is subject to change as a function of doping,  external or chemical pressure \cite{Hirschfeld_RPP_2011}. Recent theory works using a five orbital tight binding model show a near degeneracy between $d$ and $s^\pm$ pairing states in Fe-SC's as a result of the multiorbital structure of the Cooper pairs and the near nesting conditions \cite{Graser_NJP_2009, Fernandes_PRL_2013}.

Hydrostatic pressure is a clean tuning parameter that can modify the orbital overlap, the exchange interactions, and the band structure of metals. Given the near degeneracy between different pairing states, it is conceivable that the pairing symmetry of certain Fe-SC's might be tuned by pressure \cite{Tafti_NatPhys_2013, Tafti_PRB_2014, Terashima_PRB_2014, Taufour_PRB_2014, Tafti_PRB_2015, Guguchia_NatCom_2015,Wiecki_PRB_2018}. Such idea leads, in particular, to observation of a V-shaped $T-P$ phase diagram in $A$Fe$_2$As$_2$ ($A=$K, Cs, and Rb), where the transition temperature $T_{\rm c}$ decreases initially as a function of pressure, then at a critical pressure $P_{\rm c}$, it suddenly changes direction and increases \cite{Tafti_NatPhys_2013, Tafti_PRB_2014, Terashima_PRB_2014} or remains almost constant at least up to pressures $\sim 4$~GPa \cite{Taufour_PRB_2014,Grinenko_PRB_2014,Wang_PRB_2016, Wiecki_PRB_2018}.
The constancy of the Hall coefficient through $P_{\rm c}$ rules out a change in the Fermi surface \cite{Tafti_NatPhys_2013, Tafti_PRB_2014, Tafti_PRB_2015}.  In a case of KFe$_2$As$_2$ this was additionally confirmed by de Haas-van Alphen measurements, showing quantum oscillations that smoothly evolve across  $P_{\rm c}$~\cite{Terashima_PRB_2014}.  In the absence
of any sudden change in the Fermi surface across the critical
pressure $P_{\rm c}$, the resistivity and magnetic susceptibility experiments of Refs.~\onlinecite{Tafti_NatPhys_2013, Tafti_PRB_2014, Tafti_PRB_2015, Taufour_PRB_2014} detect a sudden change of the upper critical field $H_{\rm c2}(T)$, which is interpreted as an evidence of a sudden change in the structure of the superconducting energy gap across $P_{c}$.

In this paper, we report on zero-field (ZF) and longitudinal field (LF,
where a magnetic field is applied parallel to the initial muon-spin polarization) muon-spin rotation  ($\mu$SR) studies of $A$Fe$_2$As$_2$ ($A=$ K, Cs, and Rb) as a function of pressure.
In all three systems studied the magnetism was found to appear for pressures slightly below the critical one ($P_{\rm c}$), {\it i.e.} at pressure where $T_{\rm c}(P)$ changes the slope. Rather than compete \cite{Sachdev_Science_2000}, magnetism and superconductivity in $A$Fe$_2$As$_2$ are coexisting at $P\gtrsim P_{\rm c}$ pressure region.
Our findings reveal an intriguing interplay of magnetism and superconductivity in $A$Fe$_2$As$_2$, whereby the former supports (or even enhances) the latter, in contrast to the usual suppression caused by phase competition.

\begin{figure*}[t]
\includegraphics[width=0.9\linewidth]{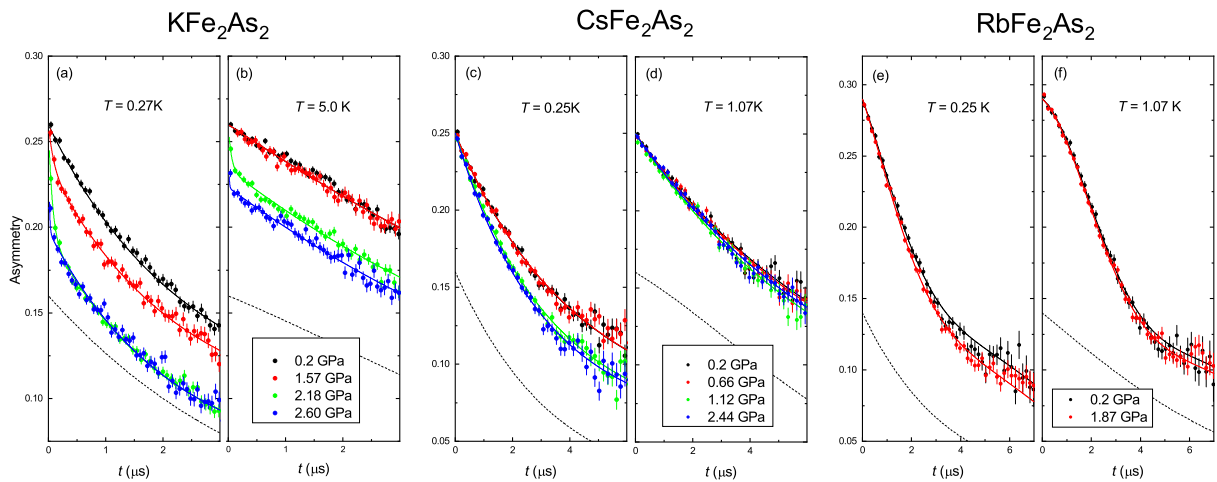}
%
\caption{Zero-field $\mu$SR time spectra measured at various pressures and temperatures:  (a) and (b) in KFe$_2$As$_2$ at $P=0.2$, 1.57, 2.18, and 2.60 ~GPa and at $T=0.27$ and 5.0~K, respectively. (c) and (d) in CsFe$_2$As$_2$ at $P=0.2$, 0.66, 1.12, and 2.44~GPa and  at $T=0.25$ and 1.07~K. (e) and (f) in RbFe$_2$As$_2$ at $P=0.2$ and 1.87~GPa and at $T=0.25$ and 1.07~K. The solid lines are fits of Eq.~\ref{eq:ZF_Asymmetry_LTF} with the sample part described by Eq.~\ref{eq:zf_K122} for KFe$_2$As$_2$ and Eq.~\ref{eq:zf_Cs122-Rb122} for CsFe$_2$As$_2$ and RbFe$_2$As$_2$, respectively. The dashed lines are the background (pressure cell) contributions. See text for details. }
 \label{fig:ZF-time-spectra}
\end{figure*}



Polycrystalline samples of KFe$_2$As$_2$, CsFe$_2$As$_2$, and RbFe$_2$As$_2$ were synthesized by conventional solid state reaction method using $A$As ($A=$ K, Cs, and Rb) and Fe$_2$As as starting materials \cite{Bukowski_PhysC_2010, Shermadini_PRB_2010}.  $A$As was prepared by reacting stoichiometric alkali metal pieces and As powders at 200$^{\rm o}$C for 4~h in an evacuated quartz tube, and Fe$_2$As was synthesized by heating Fe and As powders at 700$^{\rm o}$C for 24~h. Stoichiometric amounts of the starting materials were first thoroughly grounded, pressed into pellets, and finally sealed in a Nb tube under 1.5~atm of argon gas. The Nb tube was then sealed in an evacuated quartz tube and heated to 650$^{\rm o}$C for 36~h, with an intermediate grinding. All the sample preparation process were carried out in a glove box under high purity argon atmosphere.
The powder x-ray diffraction patterns are consistent with those reported in Ref.~\onlinecite{Rotter_AngChem_2008}.
The detailed information on the sample characterization is given in the Supplemental part, Ref.~\onlinecite{Supplemental-part}.


The pressure was generated in a double-wall piston-cylinder type of cells especially designed to perform muon-spin rotation/relaxation ($\mu$SR) experiments under pressure \cite{Khasanov_HPR_2016, Shermadini_HPR_2017}.  As a pressure transmitting medium 7373 Daphne oil was used. The maximum safely reachable pressure is $\sim 2.7$~GPa. The pressure was measured in situ by monitoring the pressure induced shift of the superconducting transition temperature of In \cite{Shermadini_HPR_2017}.


Zero-field  and longitudinal-field  $\mu$SR measurements  were  performed at the $\pi$M3 and $\mu$E1 beam lines (Paul Scherrer Institute,Villigen, Switzerland), by using the dedicated LTF and GPD spectrometers, respectively. At the LTF spectrometer, equipped with a dilution fridge cryostat, ZF-$\mu$SR experiments at ambient pressure and down to temperatures $\simeq20$~mK  were carried out. At the GPD spectrometer \cite{Khasanov_HPR_2016}, equipped with an Oxford sorption pumped He-3 cryostat (base temperature $\sim 0.24$~K), ZF and LF-$\mu$SR experiments under the pressure up to $\sim$2.7~GPa were conducted. The typical counting statistics were $\sim5-7\cdot 10^{6}$ decay positron events for each particular data point. The data were analyzed using the free software package \textsc{MUSRFIT} \cite{Suter_PhysProc_2012}.




The magnetic response of $A$Fe$_2$As$_2$ at ambient pressure was studied in set of ZF-$\mu$SR measurements at the LTF instrument.
A few representative $\mu$SR time spectra taken below and above the superconducting transition temperature for KFe$_2$As$_2$, CsFe$_2$As$_2$, and RbFe$_2$As$_2$ are shown in the Supplemental part, Ref.~\onlinecite{Supplemental-part}  and  in the Ref.~\onlinecite{Shermadini_PRB_2010}.
No sign of magnetism was detected in the ZF-$\mu$SR response of RbFe$_2$As$_2$ \cite{Shermadini_PRB_2010}. The experimental data are well described by a standard Kubo-Toyabe depolarization function \cite{Yaouanc_book_2011}, reflecting the field distribution at the muon site created by the nuclear moments.
In KFe$_2$As$_2$ and CsFe$_2$As$_2$, an exponential relaxation was found to be present in ZF-$\mu$SR signal at ambient pressure and it increases slightly with decreasing temperature thus pointing to the presence of magnetism. The absence of spontaneous oscillations indicates that there is no onset of long range magnetic order. It was also found that the relaxation occurs only in part of the sample volume ($\simeq 70$\% in a case of KFe$_2$As$_2$ and $\simeq 40$\% in a case of CsFe$_2$As$_2$), while the rest of the sample display a non-relaxing behavior, typical of a non-magnetic fraction (see the Supplemental Part, Ref.~\onlinecite{Supplemental-part}).

We think that the magnetic signal observed at the  ambient pressure in KFe$_2$As$_2$ and CsFe$_2$As$_2$ is not intrinsic for these compounds but caused by the presence of small amount of magnetic impurities in samples studied.  Our conclusions are based on the following arguments:\\
First, our longitudinal field (LF) $\mu$SR experiments of KFe$_2$As$_2$ show that at $T=0.25$~K the slow-relaxing component associated with the ambient pressure impurity magnetism [$\lambda_{Imp}(0.25{\rm ~K})\simeq 0.55 \ \mu{\rm s}^{-1}$] recovers at fields of the order of 2~mT (see Supplemental Material, Ref.~\onlinecite{Supplemental-part}). Such a behaviour is expected in a case of diluted and randomly oriented magnetic moments giving rise to weak magnetism \cite{Walstedt_PRB_1974, Yaouanc_book_2011}.\\
Second, the weak ambient pressure magnetism does not occupy the whole sample volume. Substantial part of KFe$_2$As$_2$ and CsFe$_2$As$_2$ samples remain nonmagnetic down to the lowest temperature, Ref.~\onlinecite{Supplemental-part}. \\
Third, ZF-$\mu$SR studies of the superconducting Ba$_{1-x}$K$_x$F$_2$As$_2$ with $0.5\leq x \leq 0.9$ \cite{Lotfi_Mahyari_PRB_2014}, show the absence of any kind of magnetic response. The ZF-$\mu$SR time spectra were well described by a single component exponential relaxation function with weak depolarization rate $\lambda\simeq 0.01-0.04$~$\mu{\rm s}^{-1}$. Note that these values are slightly smaller but comparable to $\lambda(T)\simeq 0.08$~$\mu{\rm s}^{-1}$ for $T\gtrsim1$~K as obtained in our study, Ref.~\onlinecite{Supplemental-part}. \\
Fourth, $\mu$SR experiments on high quality KFe$_2$As$_2$ single crystal performed by Ohishi \cite{Ohishi_private} also show a temperature independent ZF-$\mu$SR time spectra down to $T\simeq20$~mK with $\lambda\simeq0.05$~$\mu{\rm s}^{-1}$. \\
Fifth, no ambient pressure magnetism was detected for the  RbFe$_2$As$_2$ sample, Ref.~\onlinecite{Shermadini_PRB_2010} and for the second (the 'cross-checked') KFe$_2$As$_2$ sample.


\begin{figure}[htb]
\includegraphics[width=0.9\linewidth]{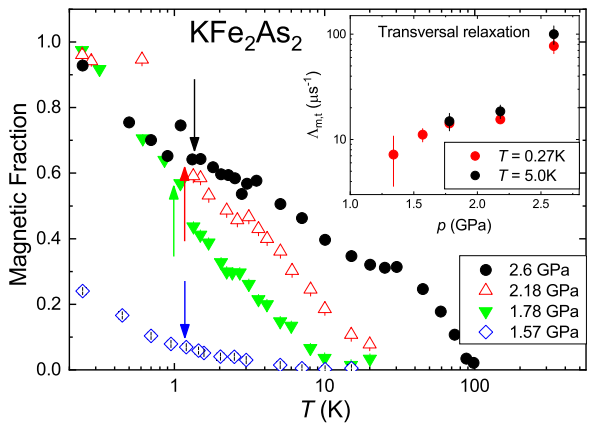}
%
\caption{ Temperature dependence of the magnetic volume fraction $f$ of KFe$_2$As$_2$ as obtained from the fit of Eq.~\ref{eq:ZF_Asymmetry_LTF} (with the sample contribution described by Eq.~\ref{eq:zf_K122}) to ZF-$\mu$SR data taken at
$P = 1.57$, 1.78, 2.18, and 2.6~GPa. Arrows indicate the position of $T_{\rm c}(P)$'s. The inset shows the pressure dependence of the fast relaxing component $\Lambda_{m,t}$. }
 \label{fig:K122_volume-fraction}
\end{figure}

As a next step, ZF-$\mu$SR experiments under a pressure were performed at the GPD instrument.  Figure~\ref{fig:ZF-time-spectra} shows ZF-$\mu$SR time spectra taken at various pressures and temperatures for all three systems studied. The build-up of an additional relaxing component, reflecting the presence of the pressure induced magnetism, is clearly visible in the spectra.

In the case of KFe$_2$As$_2$ [Figs.~\ref{fig:ZF-time-spectra}~(a) and (b)] the fast relaxing component start to develop for pressures exceeding $P\gtrsim 1.6$~GPa.  The absence of discrete frequency oscillations indicates that the magnetism is highly disordered. The amplitude of this fast relaxing  component saturates above 1.78~GPa at $T=0.27$~K and increases continuously with increasing pressure above $P\simeq$ 1.8~GPa at $T\simeq5$~K.
Pressure induced magnetism is also clearly visible in CsFe$_2$As$_2$ and RbFe$_2$As$_2$ in Figs.~\ref{fig:ZF-time-spectra}~(c) and (e) respectively.
In both systems at $T\simeq 0.25$~ K an additional relaxing contribution
starts to develop for pressures exceeding  $\sim0.7$~GPa. These additional relaxation is less pronounced in in RbFe$_2$As$_2$ compared to CsFe$_2$As$_2$ and in CsFe$_2$As$_2$ compared to KFe$_2$As$_2$.

The pressure induced magnetism observed in $A$Fe$_2$As$_2$ could arise either from static field distributions or fluctuating fields. In order to distinguish between these two possibilities, LF-$\mu$SR experiments on KFe$_2$As$_2$ were performed at $P=1.8$ and 2.6~GPa (see the Supplemental Material part, Ref.~\onlinecite{Supplemental-part}).
At both pressures studied the full muon-spin polarization was found to recover at the longitudinal field $B_{LF}\sim 0.2-0.4$~T, thus implying that the disordered pressure induced magnetism in KFe$_2$As$_2$ is static \cite{comment}.

\begin{figure*}[htb]
\includegraphics[width=0.9\linewidth]{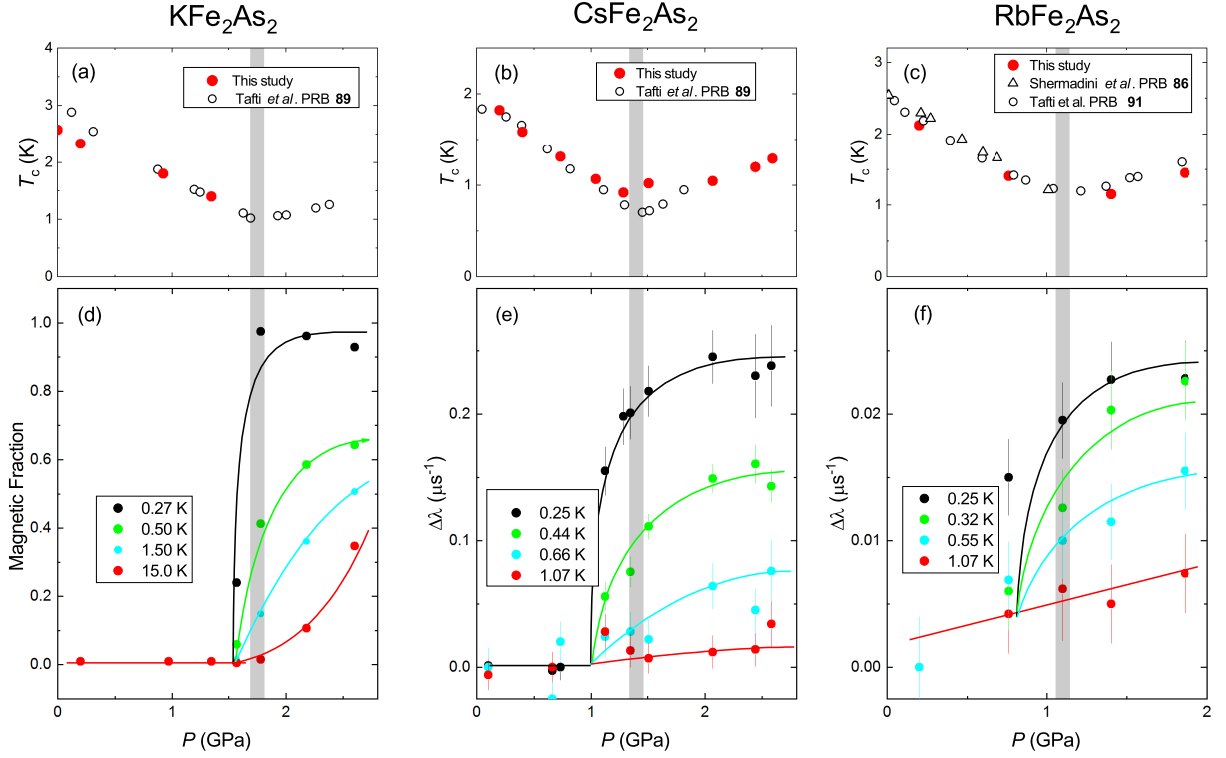}
%
\caption{(a) Dependence of the superconducting transition temperature $T_{\rm c}$ on $P$ of KFe$_2$As$_2$. The red symbols are $T_{\rm c}(P)$'s obtained in the present study in transverse-field $\mu$SR experiments \cite{Supplemental-part}. The open symbols are $T_{\rm c}(P)$ points from Ref.~\onlinecite{Tafti_PRB_2014}. (b) and (c) The same as in panel (a) but for CsFe$_2$As$_2$ and RbFe$_2$As$_2$, respectively. The open symbols in panel (c) are $T_{\rm c}(P)$ data from Refs.~\onlinecite{Tafti_PRB_2015} and \onlinecite{Shermadini_PRB_2012}. (d)  magnetic volume fraction $f$ of KFe$_2$As$_2$ on pressure at $T = 0.27$, 0.50, 1.50, and 15.0~K. (e) Dependence of the additional exponential relaxation $\Delta\lambda$ of CsFe$_2$As$_2$ on pressure for $T=0.25$, 0.44, 0.66, and 1.07~K. (f) Same is in panel (e) but for RbFe$_2$As$_2$ and at $T=0.25$, 0.32, 0.55, and 1.07~K. The grey stripes are the critical pressure regions where the transition temperature $T_{\rm c}(P)$ changes the tendency from decreasing to increasing. The solid lines in panels (c)--(f) are guides for the eye.  }
 \label{fig:Phase-diagram}
\end{figure*}

We stress that the character of this disordered static magnetism appearing under pressure is very  different from the impurity magnetism detected in KFe$_2$As$_2$ and CsFe$_2$As$_2$ at ambient pressure (see Figs.~7 and 8 in Supplemental Part, Ref.~\onlinecite{Supplemental-part}).\\
First, the impurity magnetism, which is  already present in the $\mu$SR spectra at the lowest pressure ($P \simeq 0.2$~GPa),
is clearly different from that observed in spectra taken at $P \gtrsim 1.6$~GPa for KFe$_2$As$_2$ [Figs.~\ref{fig:ZF-time-spectra}~(a) and (b)] and at $P\gtrsim 1.0$~GPa for CsFe$_2$As$_2$ [Fig.~\ref{fig:ZF-time-spectra}~(c)].\\
Second, in a case of KFe$_2$As$_2$ the relaxation is much stronger at high pressures (see the inset in Fig.~\ref{fig:K122_volume-fraction}) and the magnetism itself  extends up to much higher temperatures than that observed at ambient and low pressures.\\
Third, as will be shown later, in KFe$_2$As$_2$ for temperatures below $T \simeq 0.5$~K and pressures above $P \simeq 1.8$~GPa the high pressure magnetism occupies almost 100\% of the sample volume (see Figs.~\ref{fig:K122_volume-fraction} and \ref{fig:Phase-diagram}).\\
Finally, the similar pressure induced magnetism was found in the second (the 'cross-checked') KFe$_2$As$_2$ sample, which has no detectable amount of magnetic impurity fraction (see the Supplemental Part, Ref.~\onlinecite{Supplemental-part} for details).

The ZF high-pressure data were analyzed with the signal decomposed in contributions of the sample (s) and the pressure cell (pc):
\begin{equation}
A(t)=A_{\rm s}(0) P_{\rm s}(t)+ A_{\rm pc}(0) P_{\rm pc}(t).
 \label{eq:ZF_Asymmetry_LTF}
\end{equation}
Here $A_{\rm s}(0)$ and $A_{\rm pc}(0)$ are the initial asymmetries and $P_{\rm s}(t)$ and $P_{\rm pc}(t)$ are temperature evolutions of the muon-spin polarization belonging to the sample and the pressure cell, respectively. Note that in pressure experiments a large fraction of the muons (roughly 50\%) stops in the pressure cell walls providing a background contribution \cite{Khasanov_HPR_2016}. The pressure cell contribution to the polarization signal $P_{\rm pc}(t)$ was measured in an independent experiment.

In order to account for the strong magnetism in KFe$_2$As$_2$ the response of the sample was assumed to consist of "magnetic" and "nonmagnetic" (including an impurity contribution [$Imp(t)$] as described the Supplemental Part, Ref.~\onlinecite{Supplemental-part}) terms  and is derived as:
\begin{eqnarray}
 P_s(t)&=&
 \left[ f \left(\frac{1}{3}\;  e^{-\Lambda_{m,l}t}+\frac{2}{3}\; e^{-\Lambda_{m,t}t} \right) +(1-f)\;G_{KT}(t)\right]
 \nonumber \\
 &&\times Imp(t).
 \label{eq:zf_K122}
\end{eqnarray}
Here $f$ is the relative weight (volume) of the pressure-induced magnetic fraction. $\Lambda_{m,l}$ and $\Lambda_{m,t}$ are the exponential depolarization rates representing the longitudinal (1/3) and the transversal (2/3) relaxing components within the parts of the sample being in the magnetic state \cite{Yaouanc_book_2011}. $G_{KT}(t)=1/3+2/3 (1-\sigma_s^2t^2)e^{-\sigma_s^2t^2/2}$ is the Gaussian Kubo-Toyabe depolarization function reflecting the field distribution created by the nuclear moments, and $\sigma_s$ is the Gaussian depolarization rate caused them \cite{Yaouanc_book_2011}.  During the fit the $Imp(t)$ term was fixed to that obtained in the ZF experiments under the ambient pressure (see the Supplemental Part, Ref.~\onlinecite{Supplemental-part}). The solid lines in  Figs.~\ref{fig:ZF-time-spectra}~(a) and (d) represent the result of the fit of Eq.~\ref{eq:ZF_Asymmetry_LTF} to the ZF-$\mu$SR KFe$_2$As$_2$ data with the sample part described by Eq.~\ref{eq:zf_K122}.

Figure~\ref{fig:K122_volume-fraction} shows the dependence of the magnetic volume fraction $f$ of KFe$_2$As$_2$ on temperature for $P=1.57$, 1.78, 2.18, and 2.6~GPa. The inset shows the pressure dependence of the fast relaxing component $\Lambda_{m,t}$. Note that for $P\lesssim1.3$~GPa the fit of  Eq.~\ref{eq:zf_K122} to the experimental data result in $f$ being close to zero,  thus implying that at low pressures the KFe$_2$As$_2$ sample is "nonmagnetic" with only the impurity magnetism as observed in ZF-$\mu$SR experiment under  ambient pressure.
The data presented in Fig.~\ref{fig:K122_volume-fraction} show that with increasing pressure an increasingly large part of the sample becomes magnetic.
More important, for pressures exceeding $P \simeq 1.8$~GPa the low temperature value of $f$ reaches almost 1.0.
Given that in KFe$_2$As$_2$ bulk superconductivity is observed up to at least $P \simeq 6$~GPa \cite{Taufour_PRB_2014},
we conclude that for $P > 1.8$~GPa bulk magnetism and bulk superconductivity coexist within the whole sample volume.
Note that a similar type of bulk coexistence between the antiferromagnetic order and superconductivity was previously reported for various compounds belonging to the so-called  11 \cite{Bendele_PRL_2010, Khasanov_FeSe_PRL_2010, Bendele_PRB_2012, Kothapalli_NatCom_2016, Bohmer_PRB_2019, Holenstein_PRB_2016, Holenstein_PRL_2019}, 122 \cite{Goldman_PRB_2008, Avci_PRB_2011, Kim_PRB_2011, Kreyssig_PRB_2010, Wang_PRL_2012, Luo_PRL_2012, Nandi_PRL_2010, Marsik_PRL_2010, Christianson_PRL_2009, Pratt_PRL_2011, Wiesenmayer_PRL_2011, Bernhard_PRB_2012, Li_PRB_2012, Materne_PRB_2015, Kawasaki_PRB_2015, Sheveleva_Arxiv_2020}, 1144 \cite{Ding_PRB_2017, Budko_PRB_2018, Kreyssig_PRB_2018, Khasanov_Ni1144_arxiv_2020}, and  21311 \cite{Holenstein_Arxiv_2019} families of Fe-SC's.

The much weaker pressure-induced magnetism in CsFe$_2$As$_2$ and RbFe$_2$As$_2$ was probed via direct comparison between the low and the high-pressure data:
\begin{equation}
  P_{\rm s}^{p}(t)= P_{\rm s}^{0.2{\rm GPa}}(t) e^{-\Delta \lambda (t)}.
  \label{eq:zf_Cs122-Rb122}
\end{equation}
Here $P_{\rm s}^{0.2{\rm GPa}}(t)$ is the time evolution of the muon-spin polarization measured at $P\simeq 0.2$~GPa (low pressure) and $\Delta\lambda$ is the additional exponential relaxation caused by pressure.
The solid lines in  Figs.~\ref{fig:ZF-time-spectra}~(c)--(f) represent the result of the fit of Eq.~\ref{eq:ZF_Asymmetry_LTF} to the ZF-$\mu$SR CsFe$_2$As$_2$ and RbFe$_2$As$_2$ data with the sample part described by  Eq.~\ref{eq:zf_Cs122-Rb122}.

Figure~\ref{fig:Phase-diagram} summarizes the results of our ZF-$\mu$SR study on a $f-P$  and $\Delta\lambda-P$ phase diagrams. For comparison we have also plotted the $T_{\rm c}(P)$ curves as they obtained in resistivity, magnetization and transverse-field $\mu$SR experiments \cite{Tafti_PRB_2014, Tafti_PRB_2015, Shermadini_PRB_2012, Supplemental-part}.
The grey stripes indicate the critical pressure regions: $P_{\rm c}\simeq 1.75$, 1.4, and 1.1~GPa for KFe$_2$As$_2$, CsFe$_2$As$_2$, and RbFe$_2$As$_2$, respectively.
In KFe$_2$As$_2$ the magnetic volume fraction increases rapidly from nearly zero to $\sim$~100~\% between 1.5~GPa and 1.8~GPa, at least at the lowest temperature $T\simeq 0.27$~K [Fig.~\ref{fig:Phase-diagram}~(d)]. In CsFe$_2$As$_2$ and RbFe$_2$As$_2$ [Figs.~\ref{fig:Phase-diagram}~(e) and (f)] an additional exponential relaxation $\Delta\lambda$ goes to saturation for $P$ exceeding $\simeq 1.5$ and $\simeq1.2$~GPa, respectively.
It is remarkable that in all three systems the magnetism develops {\it closely} to the $P_{\rm c}$, {\it i.e} to the pressure where the $T_{\rm c}(P)$ curve acquires the positive slope.

It appears, therefore that the disordered magnetism (which is static at least in the case for KFe$_2$As$_2$, Ref.~\onlinecite{comment}) supports  (or even enhances) superconductivity in all three systems studied.
This is unusual, as static magnetic order typically tends to compete with superconductivity.
Note, however, that the magnetism we detect appears to be highly disordered and as such does not break translational symmetry on a sufficient length scale
to cause the usual Fermi-surface reconstruction that is detrimental to superconductivity.
Indeed, no change is detected in the Fermi surface across $P_{\rm c}$ \cite{Tafti_NatPhys_2013, Tafti_PRB_2014, Tafti_PRB_2015, Terashima_PRB_2014, Taufour_PRB_2014}.
The cooperative nature of the interplay between magnetism and superconductivity is also reflected by the fact that the magnetism enhances significantly below $T_{\rm c}$ in a case of KFe$_2$As$_2$ [Fig.~\ref{fig:K122_volume-fraction} and Fig.~\ref{fig:Phase-diagram}~(a)] and develops just below $T_{\rm c}$ in a case of CsFe$_2$As$_2$ and RbFe$_2$As$_2$ [Figs.~\ref{fig:Phase-diagram}~(b) and (d)].

This work was performed at the Swiss Muon Source (S$\mu$S), Paul Scherrer Institute (PSI, Switzerland).
The authors are grateful to L.~Taillefer for initiating the work and for helpful scientific discussions at the beginning of the project.  A.~Amato is acknowledged  for help during the $\mu$SR experiments. R.K. acknowledges supporting discussions with J.~Mesot.

\end{document}